\documentclass[12pt,letterpaper,american]{article}
\usepackage[T1]{fontenc}
\usepackage[latin1]{inputenc}
\usepackage{graphicx}

\makeatletter

\usepackage{setspace}
\usepackage{graphicx}
\usepackage[scanall]{psfrag}

\usepackage{babel}
\makeatother
\begin{document}

\title{Global Potential Energy Minima of (H$_{2}$O)$_{n}$ Clusters on
Graphite}

\author{B. S. González, J.~Hern\'{a}ndez-Rojas, J.~Bret\'{o}n,\\ and J.~M.~Gomez
Llorente%
\thanks{Correponding author. \emph{E-mail address}: jmgomez@ull.es %
}\\
 Departamento de F\'{\i}sica Fundamental II\\
 Universidad de La Laguna, 38205 Tenerife, Spain}

\maketitle
\begin{abstract}
Likely candidates for the global potential energy minima of (H$_{2}$O)$_{n}$
clusters with $n\leq21$ on the (0001)-surface of graphite are found
using basin-hopping global optimization. The potential energy surfaces
are constructed using the TIP4P intermolecular potentials for the
water molecules (the TIP3P is also explored as a secondary choice),
a Lennard-Jones water-graphite potential, and a water-graphite polarization
potential that is built from classical electrostatic image methods
and takes into account both the perpendicular and parallel electric
polarizations of graphite. This potential energy surface produces
a rather hydrophobic water-graphite interaction. As a consequence,
the water component of the lowest graphite-(H$_{2}$O)$_{n}$ minima
is quite closely related to low-lying minima of the corresponding
TIP4P (H$_{2}$O)$_{n}$ clusters. In about half of the cases the
geometrical substructure of the water molecules in the graphite-(H$_{2}$O)$_{n}$
global minimum coincides with that of the corresponding free water
cluster. Exceptions occur when the interaction with graphite induces
a change in geometry. A comparison of our results with available theoretical
and experimental data is performed. 
\end{abstract}
\doublespacing

\section{Introduction}

\label{sec1}The interaction of carbonaceous materials such as fullerenes,
carbon nanotubes and graphite with atoms and molecules share many
properties. In particular, in this work we will be concerned with
the interaction between water and graphite. A deep understanding of
the features and properties of this interaction is of particular interest
in technological applications, such as those related with the use
of water as a lubricant for graphite \cite{lancaster,zaidi}, and,
more indirectly, in the behavior of water at the nanometer scales
when material related to graphite, such as carbon nanotubes, are present.
Water-graphite interaction is also relevant in the design of corrosion-free
combustion chambers and rocket nozzles, since water is a universal
combustion product and graphite is an important surface material because
of its chemical inertness under extreme conditions \cite{xu}. Other
fields benefiting from this knowledge include the environmental sciences
\cite{popo} and astrophysics \cite{draine}, since graphite is a
good candidate for the composition of nano particles and dust grains.

Despite the natural abundance of water and graphite, relatively few
experimental data are available for their interaction. Studies at
low temperature ($T=85$ K) and low coverage using temperature programed
desorption and vibrational high resolution electron energy loss spectroscopy
have shown that water is adsorbed non dissociatively on the graphite
surface forming hydrogen bonded aggregates with a two dimensional
structure that changes into a three dimensional one upon warming \cite{chaca}.
The water arrangement for the two-dimensional structure is unknown,
as is also unknown the role played by small water clusters in the
growth of these structures. 

Experimental information about the water-graphite binding energies
and structural aspects is currently lacking even for the water-monomer
adsorption. To our knowledge, there are only the early water-graphite
binding energy by Kieslev \emph{et al.} (15.0 kJ/mol) \cite{avgul}
and the more recent association energy reported by Kasemo \emph{et
al}. \cite{chaca}.

In the last few years some results from theoretical calculations have
been made available. Some of these studies are concerned with macroscopic
features of the water-graphite interaction. In this group we can include
the work by Werder \emph{et al.} \cite{werder}, who fit an interaction
potential form to experimental data for the contact angle of water
nanodroplets on graphite surfaces. A similar scheme is used by Pertsin
\emph{et al}. to simulate lubricant properties from a water-graphite
interaction that was previously fitted to ab-initio \cite{pertsin1}
and empirical data \cite{pertsin2}. Finally Gatica \emph{et al}.
have used empirical water-graphite potentials to look for a wetting
transition \cite{gatica}.

\emph{Ab initio} calculations have been recently reported. By using
second-order Möller-Plesset perturbation theory, Feller \emph{et al}.
\cite{feller} have provided the interaction energy between a water
molecule and acenes as large as C$_{96}$H$_{24}$; the value of 24
kJ/mol that was obtained for this energy seems to be unphysically
high \cite{karapetian}. This conclusion is confirmed by the recent
theoretical calculations by Sudiarta and Geldart \cite{sudiarta}.
Using the same Möller-Plesset scheme for a water molecule on both
hydrogen and fluorine terminated acenes, these authors demonstrated
the important contribution of this boundary to the water-acene binding
energy. After removing this effect and extrapolating to an infinite
graphene they reported a value of 10.2 kJ/mol. Density functional
theory (DFT) total energy calculations were performed by Cabrera Sanfélix
\emph{et al}. \cite{cabrera} to study structural aspects of water
layers on graphite. Absolute binding energies were not provided and
no global energy minimization was done. These tasks are performed
in later calculations using DFT tight-binding methods complemented
with empirical van der Waals force corrections (DFTB-D) \cite{xu,lin}.
In these works, clusters with up to 6 water molecules on graphite are studied and
the optimal structures and the binding and association energies were
provided. Besides, these results are compared with those obtained
with integrated ONIOM (ab-intio B3LYP:DFTB-D$+$semiempirical PM3)
methods \cite{xu}. In these studies graphite is represented by up
to three-layer acenes. In all these DFT studies the acene boundary
effects are not removed.

A full empirical approach has been followed by Karapetian \emph{et
al}. \cite{karapetian} by using the Dang-Chang model for the water-water
interaction and a polarizable potential model for the water-graphene
interaction that includes dispersion-repulsion contributions by means
of Lennard-Jones pairwise interactions. In this potential the polarization
term is built by associating an isotropic polarizable center with
each carbon atom. The interaction between these centers, when polarized,
is neglected. Again, only clusters with up to 6 water molecules are
considered.

A similar empirical approach shall be followed in our present work.
Because of its ability to reproduce the structure of water clusters,
we will use the TIP4P model for the water-water interaction \cite{jorge}.
In order to analyze dependence of our results on this choice, we shall
also consider the TIP3P water-water interaction model \cite{jorge}.
The water-graphite interaction analytic model shall include a dispersion-repulsion
term, built from the sum of infinite Lennard-Jones pairwise interactions
using the Steele method, and a polarization contribution which shall
be built using electrostatic image methods that take into account
the anisotropic response of graphite. We shall provide likely candidates
for the global potential energy minima of graphite-(H$_{2}$O)$_{n}$
clusters with $n\leq21$. We shall employ basin-hopping global optimization
to identify these global minima. From the structure and energetics
of these minima we shall elucidate about the hydrophobic nature of
the water-graphite interaction at the lowest temperatures, and the
dependence of the results on the potential model.

This paper is organized as follows. In Section \ref{sec2} we discuss
our expression for the potential energy surface as a sum of Coulomb,
dispersion-repulsion, and polarization contributions. In Section \ref{sec3}
we present likely candidates for the cluster global potential energy
minima together with their association and binding energies. Here
we shall compare our values with the available data. Finally, Section
\ref{sec4} summarizes our conclusions.

\section{The Potential Energy Function}

\label{sec2}The closed-shell electronic structure of both graphite
and water makes an empirical approach to the potential energy surface
(PES) for the water-graphite and water-water interactions particularly
attractive. We write the potential energy of a graphite-(H$_{2}$O)$_{n}$
cluster as a sum of two contributions \begin{equation}
V=V_{{\rm ww}}+V_{{\rm wg}},\label{eq:1}\end{equation}
where $V_{{\rm ww}}$ is the sum of pairwise water-water interactions,
and $V_{{\rm wg}}$ is the water-graphite term. For the water-water
interaction we have chosen the TIP4P form as a primary choice, but
we will also consider the TIP3P model. These models describe each
water molecule as a rigid body with two positive charges on the hydrogen
atoms and a balancing negative charge either close to the oxygen atom
(TIP4P) or just at the oxygen atom (TIP3P), together with a dispersion-repulsion
center on the oxygen atom. Hence, $V_{{\rm ww}}$ is a sum of pairwise
additive Coulomb and Lennard-Jones terms. These models have been used
in the study of homogeneous water clusters \cite{wales0,kabrede,hartke1},
water clusters containing metallic cations \cite{briesta,hartke2},
and water-C$_{60}$ clusters \cite{rojas}.

The water-graphite interaction is written as \begin{equation}
V_{{\rm wg}}=V_{{\rm dr}}+V_{{\rm pol}},\label{eq:2}\end{equation}
where $V_{{\rm dr}}$ is a sum of pairwise dispersion-repulsion terms
between the oxygen and the carbon atoms. Each of these terms is expressed
as a Lennard-Jones potential, whose parameters were obtained using
the standard Lorentz-Berthelot combination rules from the corresponding
parameters for the oxygen-oxygen and carbon-carbon interactions in
TIP4P and TIP3P water and Steele \cite{steele} graphene-graphene
potentials, respectively. Specifically, we used the values $\varepsilon_{{\rm CO}}=0.389$
kJ/mol and $\sigma_{{\rm CO}}=3.28$ $\textrm{Å}$ for the TIP4P,
and $\varepsilon_{{\rm CO}}=0.385$ kJ/mol and $\sigma_{{\rm CO}}=3.28$
$\textrm{Å}$ for the TIP3P, which are similar to those derived by
Werder \emph{et al.} \cite{werder} to fit the contact angle for a
water droplet on a graphene surface. An analytic form for $V_{{\rm dr}}$
can be obtained using Steele summation method \cite{steele,ocasio}
over the graphite periodic structure by writing the interaction, $U_{{\rm dr}}$,
of a dispersion center at the point $(x,y,z)$ with a graphite layer
located at the surface $z=0$ (the origin of the reference frame is
chosen at the center of a carbon hexagon), as a Fourier series, i.e.
\begin{equation}
U_{{\rm dr}}(x,y,z)=U_{0}(z)+\sum_{l>0}U_{l}(z)f_{l}(x,y)\label{eq:3}\end{equation}
We have checked that the contribution to this expansion from terms
with $l>1$ is negligible. Up to $l=1$, we have \begin{equation}
U_{0}(z)=\frac{8\pi\varepsilon_{{\rm CO}}\sigma_{{\rm CO}}^{2}}{\sqrt{3}a_{0}^{2}}\left[\frac{2}{5}\left(\frac{\sigma_{{\rm CO}}}{z}\right)^{10}-\left(\frac{\sigma_{{\rm CO}}}{z}\right)^{4}\right],\label{eq:4}\end{equation}
\begin{equation}
U_{1}(z)=\frac{8\pi\varepsilon_{{\rm CO}}\sigma_{{\rm CO}}^{2}}{\sqrt{3}a_{0}^{2}}\left[\frac{1}{60}\left(\frac{2\pi\sigma_{{\rm CO}}^{2}}{\sqrt{3}a_{0}z}\right)^{5}K_{5}(\frac{4\pi z}{\sqrt{3}a_{0}})-\left(\frac{2\pi\sigma_{{\rm CO}}^{2}}{\sqrt{3}a_{0}z}\right)^{2}K_{2}(\frac{4\pi z}{\sqrt{3}a_{0}})\right],\label{eq:5}\end{equation}
and\begin{equation}
f_{1}(x,y)=-2\left\{ \cos\left[\frac{2\pi}{a_{0}}\left(x+\frac{y}{\sqrt{3}}\right)\right]+\cos\left[\frac{2\pi}{a_{0}}\left(x-\frac{y}{\sqrt{3}}\right)\right]+\cos\left[\frac{4\pi y}{\sqrt{3}a_{0}}\right]\right\} ,\label{eq:6}\end{equation}
where $a_{0}=1.42$ $\textrm{\AA}$ is the C-C distance in the graphite
layer, $K_{m}(z)$ are the modified Bessel function of $m^{\mbox{th}}$
order, and $f_{1}(x,y)$ is the first corrugation function. The total
dispersion-repulsion interaction is obtained as a sum of $U_{{\rm dr}}$
terms over each graphite layer. We have obtained well converged values
by including the $U_{0}$ contribution from the two upper layers and
the first corrugation of the first layer. 

In Eq.~(\ref{eq:2}), $V_{{\rm pol}}$ includes the energy associated
with the polarization of graphite due to the electric field of all
the water point charges. This many-body interaction, which will turn
out to be smaller than $V_{{\rm dr}}$, was evaluated using a continuum
representation of graphite. We will evaluate two contributions to
$V_{{\rm pol}}$,\begin{equation}
V_{{\rm pol}}=V_{\parallel}+V_{\perp},\label{eq:7}\end{equation}
each one associated, respectively, with the response of graphite to
the electric field component parallel and perpendicular to the graphite
surface. For the first one, $V_{\parallel}$, we will assume that
graphite behaves as a classical conductor, which allows us to make
use of the image charge method to obtain (in Gaussian units)\begin{equation}
V_{\parallel}=-\sum_{i}\frac{q_{i}^{2}}{4z_{i}}-\frac{1}{2}\sum_{i\neq j}\frac{q_{i}q_{j}}{r_{ij}^{\prime}},\label{eq:8}\end{equation}
where $q_{i}$ is each of the water electric point charges and $r_{ij}^{\prime}$
is the distance between the charge $q_{i}$ and the image of the charge
$q_{j}$, i.e., $r_{ij}^{\prime}=[(x_{i}-x_{j})²+(y_{i}-y_{j})²+(z_{i}+z_{j})²]^{1/2}$. 

In order to evaluate $V_{\perp}$, we will assume that a graphite
layer (at $z=0$) has, in reciprocal space, a surface polarizability
density $\alpha_{\perp}(k_{x}k_{y})$ such that when an electric field
depending on the surface point and perpendicular to the layer, $E_{\perp}(x,y)=\frac{1}{2\pi}\int\mbox{d}k_{x}\mbox{d}k_{y}e^{-\mbox{i}(k_{x}x+k_{y}y)}\mathcal{E}_{\perp}(k_{x},k_{y})$,
is applied, an electric dipole density, $I_{\perp}(x,y)=\frac{1}{2\pi}\int\mbox{d}k_{x}\mbox{d}k_{y}e^{-\mbox{i}(k_{x}x+k_{y}y)}\mathcal{I}_{\perp}(k_{x},k_{y})$,
is induced on that layer, with $\mathcal{I}_{\perp}(k_{x}k_{y})=\alpha_{\perp}(k_{x}k_{y})\mathcal{E}_{\perp}(k_{x},k_{y})$.
If we now neglect the dependence of $\alpha_{\perp}$ on $k_{x}$
and $k_{y}$ (which is a valid approximation if $E_{\perp}(x,y)$
depends smoothly enough on the surface point), then we would have
$I_{\perp}(x,y)=\alpha_{\perp}E_{\perp}(x,y)$, with $\alpha_{\perp}=\alpha_{\perp}(0,0)$
being the electric polarizability in a uniform electric field perpendicular
to the layer. In this way we can calculate the dipole density induced
in the graphite layer by an electric charge $q_{i}$ at the point
$(x_{i},y_{i},z_{i})$, and from this dipole density we can evaluate
its electric field and the electric force between the polarized layer
and that charge. One readily shows that the electric field due to
the polarized surface in the half-space of the charge ($z>0$) is
equal to the electric field induced by an image dipole $p_{i}=-2\pi\alpha_{\perp}q_{i}$
at the point $(x_{i},y_{i},-z_{i})$ and direction parallel to the
$z$ axis. This result can be generalized additively to the case of
several electric point charges, all of them located in the space region
$z>0$. From the corresponding image dipoles we can obtain their electric
force on each charge, and from here the interaction potential $V_{\perp}$,
namely\begin{equation}
V_{\perp}=-\sum_{i}\frac{2\pi\alpha_{\perp}q_{i}^{2}}{8z_{i}^{2}}-\frac{1}{2}\sum_{i\neq j}\frac{2\pi\alpha_{\perp}q_{i}q_{j}(z_{i}+z_{j})}{r_{ij}^{\prime3}}.\label{eq:9}\end{equation}

The use of expression (\ref{eq:9}) requires the knowledge of $\alpha_{\perp}$.
The value of this polarizability density may be estimated from $\varepsilon_{\perp}$,
the relative electric permittivity of graphite for applied electric
fields perpendicular to the (0001) surface, whose value is $\varepsilon_{\perp}=5.75$.
Following Hannay's alternative derivation of the Clausius-Mossotti
equation \cite{hannay}, we shall require the form of the diverging
term in the expression of the electric field induced by a uniform
surface dipole density of magnitude $I_{\perp}$ and direction perpendicular
to the surface. The magnitude of this term is readily found to be
$-4\pi I_{\perp}\delta(z)$. By space averaging it, we obtain the local electric
field on each graphite layer $E_{{\rm local}}=E+4\pi I_{\perp}/d$,
where $d=3.35$ $\textrm{\AA}$ is the layer-to-layer distance and
$E$ is the magnitude of the macroscopic electric field in the medium.
Then from the relation $I_{\perp}=\alpha_{\perp}E_{{\rm local}}$, and the
expression $P=\frac{(\varepsilon_{\perp}-1)}{4\pi}E$ relating the
volume polarization density $P=I_{\perp}/d$ and $E$, we arrive at the desired
result \begin{equation}
\alpha_{\perp}=\frac{d(\varepsilon_{\perp}-1)}{4\pi\varepsilon_{\perp}}.\label{eq:10}\end{equation}
We obtain by this procedure the value $\alpha_{\perp}=0.220$ $\textrm{\AA}$. 

Being consistent with our metallic assumption for $V_{\parallel}$,
in the evaluation of the polarization contribution to the water-graphite
interaction potential we shall assume total screening of the electric
field by the external graphite surface. Therefore, only the most external
graphite layer shall be considered in this evaluation. 

We have also checked the relevance of the McLachlan substrate mediated
interaction \cite{bruch} between the water molecules in the presence
of the conducting graphite layer and found it to be negligible ($\sim0.03\%$
of the total interaction energy for the water dimer on graphite);
therefore, we shall not include this term in our potential energy.

As mentioned in the Introduction, Karapetian \emph{et al}. \cite{karapetian}
have proposed a different model potential for the polarization contribution
to the water-graphite interaction energy. These authors locate an
isotropic polarizable center at each carbon atom and calculate the
polarization energy as a sum of the contributions from each center.
This model does not take into account the collective properties of
the delocalized $\pi$ electrons and neglects completely all screening
effects among the induced dipole moments. We have estimated that these
defects of this model lead to an overestimation of the polarization
energy by a factor of two for the water monomer.

\section{Global Potential Energy Minima}

\label{sec3}Likely candidates for the global potential energy minima
of graphite-(H$_{2}$O)$_{n}$ clusters with $n\leq21$ were located
using the basin-hopping scheme \cite{wales}, which corresponds to
the `Monte Carlo plus energy minimization' approach of Li and Scheraga
\cite{li}. This method has been used successfully for both neutral
\cite{wales} and charged atomic and molecular clusters \cite{rojas,wales2,wales3,rojas1,rojas2},
along with many other applications \cite{wales4}. In the size range
considered here the global optimization problem is relatively straightforward,
but somewhat more costly than in the water-C$_{60}$ clusters \cite{rojas}.
The global minimum is generally found in fewer than $7\times10^{4}$
basin-hopping steps, independent of the random starting geometry.
In some cases, starting out from the (H$_{2}$O)$_{n}$ global potential
minimum, the corresponding global minimum for graphite-(H$_{2}$O)$_{n}$
is found even faster.

For graphite-(H$_{2}$O)$_{n}$ clusters, association energies, $\Delta E_{{\rm a}}$,
are defined for the process \begin{equation}
\mbox{graphite}+n\textrm{H}_{2}\textrm{O}=\mbox{graphite-}(\textrm{H}_{2}\textrm{O})_{n};\qquad-\Delta E_{{\rm a}}.\label{r1}\end{equation}
 We also define the water binding energy, $\Delta E_{{\rm b}}$, as
the difference between the association energies of graphite-(H$_{2}$O)$_{n}$
and (H$_{2}$O)$_{n}$, i.e. \begin{equation}
\mbox{graphite}+(\textrm{H}_{2}\textrm{O})_{n}=\mbox{graphite-}(\textrm{H}_{2}\textrm{O})_{n};\qquad-\Delta E_{{\rm b}}.\label{r2}\end{equation}
 The clusters in these expressions are assumed to be in their global
minimum. The structures and association energies employed here for
the global minima of (H$_{2}$O)$_{n}$ coincide precisely with those
obtained by Wales and Hodges \cite{wales0} and Kabrede and Hentschke
\cite{kabrede}.%
\begin{figure}
\psfrag{n}[tc][tc]{\large $n$}
\psfrag{(a)}[tc][tc]{\large $(a)$}
\psfrag{(b)}[tc][tc]{\large $(b)$}
\psfrag{1}[tc][tc]{$1$}
\psfrag{2}[tc][tc]{$2$}
\psfrag{3}[tc][tc]{$3$}
\psfrag{4}[tc][tc]{$4$}
\psfrag{5}[tc][tc]{$5$}
\psfrag{6}[tc][tc]{$6$}
\psfrag{0}[tc][tc]{$0$}
\psfrag{3}[tc][tc]{$3$}
\psfrag{6}[tc][tc]{$6$}
\psfrag{9}[tc][tc]{$9$}
\psfrag{12}[tc][tc]{$12$}
\psfrag{10}[tc][tc]{$10$}
\psfrag{20}[tc][tc]{$20$}
\psfrag{30}[tc][tc]{$30$}
\psfrag{40}[tc][tc]{$40$}
\psfrag{Energy}[bl][bl]{\hspace*{-1.5cm} Energy (kJ/mol)}
\includegraphics[width=8.25cm,keepaspectratio]{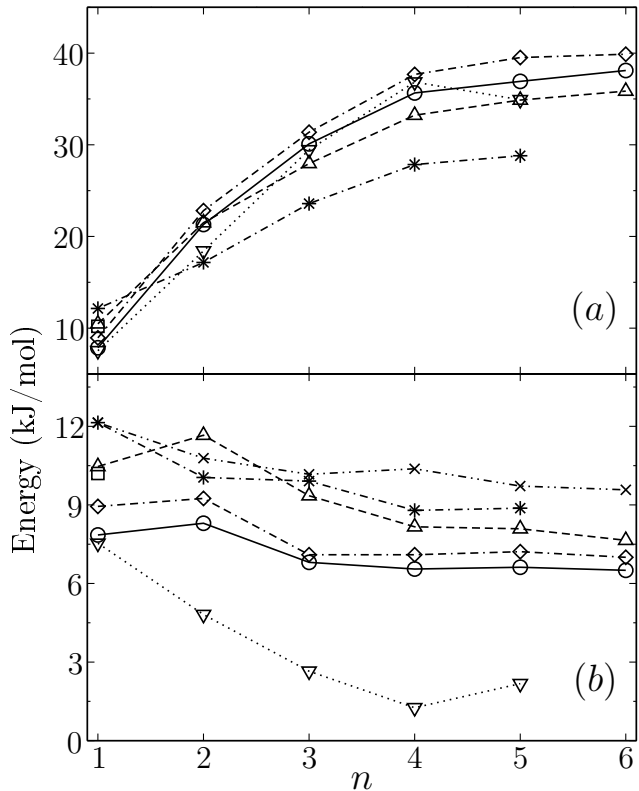}

\caption{\label{f1} Association, $\Delta E_{{\rm a}}/n$ (a), and binding,
$\Delta E_{{\rm b}}$ (b), energies per water molecule, for the global
minima of water-graphene clusters. Our results: circles for TIP4P
and diamonds for TIP3P. Other calculations: DFTB-D \cite{lin} (crosses);
DFTB-D \cite{xu} (asterisks); empirical \cite{karapetian} (up triangles);
ONIOM \cite{xu} (down triangles); Möller-Plesset \cite{sudiarta}
(square, only for $n=1$).}
\end{figure}

For comparison with the available data, we plot in Fig.~\ref{f1}
the association and binding energies defined in (\ref{r1}) and (\ref{r2})
as a function of the number of water molecules, $n\leq6$, for water
on graphene. Our energies for the TIP3P model are somewhat higher
than those for the TIP4P. With the exception of the ONIOM data, our
binding and association energies are systematically lower and higher,
respectively, than the \emph{ab initio} and other empirical potential
values. However, it is well known that DFT methods tend to overestimate
the binding energies. Furthermore, the acene-boundary effects discussed
in the Introduction have not been removed in the \emph{ab initio}
calculations, except in Sidarta and Geldart's binding energy for the
water monomer. It is also relevant to remind here the overestimation
of the polarization energy that takes place in the empirical potential
model by Karapetian \emph{et al}., as mentioned at the end of Section
\ref{sec2}. Taking all these facts into account, we can conclude
that the binding energies provided by our interaction model are quite
reliable.

Associations energies for $n\geq2$ are dominated by the water-water
interaction. As far as the empirical potential data are concerned,
differences in these association energies are a consequence of the
different water-water model interaction used in each case. The lower
\emph{ab initio} values reported by Xu \emph{et al}. can be attributed
to the poor description of the water-water interaction in their DFTB-D
scheme.

The structures of the lowest minima obtained for graphene-(H$_{2}$O)$_{n}$
with $1<n\leq6$ coincide with those provided by the other available
calculations and with the structures that we will obtain later for
the graphite-(H$_{2}$O)$_{n}$ clusters. For $n=1$, our calculation,
as the one done by Karapetian \emph{et al}., provides an H$_{2}$O
molecule with an OH bond pointing towards the graphene and the oxygen
atom just over the center of a carbon hexagon ring. On the other hand,
the \emph{ab initio} data provide a two-legged conformation with the
two OH bonds pointing towards graphene in a symmetric way. This might
be understood as the result of quantum zero-point-energy effects,
since the difference in energy between the two geometries may be of
the order of the zero point energy.

\begin{figure}
\psfrag{n}[tc][tc]{\large \vspace*{-2.5cm} $n$}
\psfrag{(a)}[tc][tc]{\large $(a)$}
\psfrag{(b)}[tc][tc]{\large $(b)$}
\psfrag{1}[tc][tc]{$1$}
\psfrag{3}[tc][tc]{$3$}
\psfrag{5}[tc][tc]{$5$}
\psfrag{7}[tc][tc]{$7$}
\psfrag{9}[tc][tc]{$9$}
\psfrag{11}[tc][tc]{$11$}
\psfrag{13}[tc][tc]{$13$}
\psfrag{15}[tc][tc]{$15$}
\psfrag{17}[tc][tc]{$17$}
\psfrag{19}[tc][tc]{$19$}
\psfrag{21}[tc][tc]{$21$}
\psfrag{0}[tc][tc]{$0$}
\psfrag{20}[tc][tc]{$20$}
\psfrag{40}[tc][tc]{$40$}
\psfrag{60}[tc][tc]{$60$}
\psfrag{80}[tc][tc]{$80$}
\psfrag{0}[tc][tc]{$0$}
\psfrag{15}[tc][tc]{$15$}
\psfrag{30}[tc][tc]{$30$}
\psfrag{45}[tc][tc]{$45$}
\psfrag{60}[tc][tc]{$60$}
\psfrag{75}[tc][tc]{$75$}
\psfrag{90}[tc][tc]{$90$}
\psfrag{Energy}[bl][bl]{\hspace*{-1.5cm} Energy (kJ/mol)}

\includegraphics[width=8.25cm,keepaspectratio]{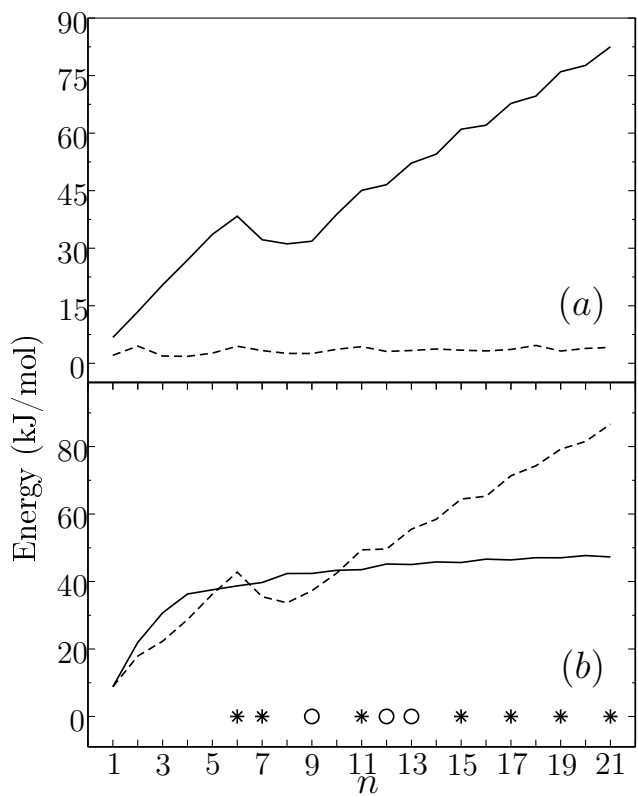}

\caption{\label{f2} (a) Polarization $V_{{\rm pol}}$ (dashed line) and dispersion-repulsion
$V_{{\rm dr}}$ (full line) contributions to the potential energy
of global minimum graphite-(H$_{2}$O)$_{n}$ clusters. (b) The corresponding
association energies per water molecule, $\Delta E_{{\rm a}}/n$ (full
line), and binding energies, $\Delta E_{{\rm b}}$ (dashed line).
Global minima in which the structure of the (H$_{2}$O)$_{n}$ moiety
differs from the global minimum of the corresponding TIP4P (H$_{2}$O)$_{n}$
cluster are marked according to the discussion in the text.}
\end{figure}

The association ($\Delta E_{{\rm a}}/n$) and binding energies ($\Delta E_{{\rm b}}$)
for the full graphite-(H$_{2}$O)$_{n}$ clusters, which have been
calculated as described in Section \ref{sec2} with the TIP4P water-water
interaction, are plotted in Fig. \ref{f2}. %
We also include in Fig.~\ref{f2}(a) the values of the polarization
energy $V_{{\rm pol}}$ and water-graphite dispersion-repulsion energy
$V_{{\rm dr}}$, as defined in Section \ref{sec2}, for the cluster
global minima. The term $V_{{\rm pol}}$ oscillates with $n$ around
an average value of $\overline{V}_{{\rm pol}}=3.33$\,kJ/mol; the
two contributions to $V_{{\rm pol}}$, $V_{\parallel}$ and $V_{\perp}$,
are similar in magnitude with $V_{\parallel}$ somewhat larger than
$V_{\perp}$. The term $V_{{\rm dr}}$ fluctuates also around a slowly
growing average as the number of water molecules close to the graphite
surface increases. On average, each of these water molecules contributes
about $7.26$\,kJ/mol to $V_{{\rm dr}}$. The water-graphite binding
energies correspond quite closely to the sum of $V_{{\rm pol}}$ and
$V_{{\rm dr}}$, while the association energies are dominated by the
water-water interaction. The average value of the association energy
per molecule in homogeneous TIP4P $(\textrm{H}_{2}\textrm{O})_{n}$
clusters with $6\leq n\leq21$ is $\sim42$\,kJ/mol \cite{wales0,kabrede}.
For water cluster on graphite the corresponding value turns out to
be $44.6$\,kJ/mol, which is comparable with the experimental value
of $43.4\pm2.9$\,kJ/mol \cite{chaca}. Any of these values corresponds
to the binding energy of a water molecule in a water cluster, and
it is much larger than the energy for binding a water molecule onto
the graphite surface. This energy balance would support an hydrophobic
nature of the water-graphite interaction.

\begin{figure}
\includegraphics[width=8.25cm,keepaspectratio]{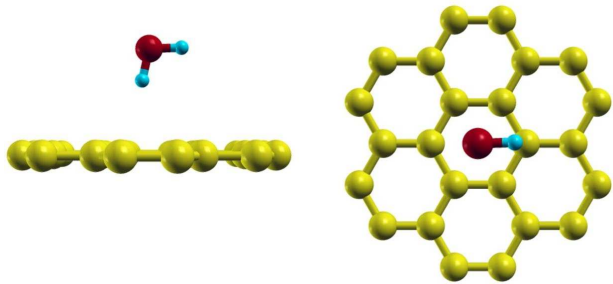}

\caption{\label{f3} Two views of the global minimum obtained for graphite-(H$_{2}$O).
This figure, and those that follow, were prepared using the program
XCrysDen \cite{xcrys}.}
\end{figure}

For $n=1$ we obtain a binding energy $\Delta E_{{\rm a}}=\Delta E_{{\rm b}}=8.81$\,kJ/mol.
This value is somewhat larger than our TIP4P binding energy for a
water molecule on graphene ($\Delta E_{{\rm b}}=7.6$\,kJ/mol), and
for the corresponding C$_{60}$-(H$_{2}$O) cluster ($\Delta E_{{\rm b}}=6.31$\,kJ/mol)
\cite{rojas}. Thus, the numbers provided above for the binding energies
of these three compounds are, at least, physically consistent. In
the present case, the contribution of the polarization energy to the
graphite-H$_{2}$O binding energy is $2.08$\,kJ/mol; the corresponding
value in the C$_{60}$-(H$_{2}$O) cluster was $2.32$\,kJ/mol \cite{rojas}.
This larger value is a consequence of the important small-size quantum
effects that make the polarizability of the C$_{60}$ molecule significantly
larger than that of a conducting sphere with the geometrical C$_{60}$
radius \cite{tomanek}. The polarization energy is responsible for
orienting the H$_{2}$O molecule with an OH bond pointing towards
the graphite surface and the oxygen atom just over the center of a
hexagonal carbon ring (Fig.~\ref{f3}). %

The angle between the water C$_{2}$ symmetry axis and our $z$-axis
in the $n=1$ global minimum is 39.8 degrees, practically identical to the
corresponding value in the C$_{60}$-(H$_{2}$O) cluster (40.4 degrees)
\cite{rojas} and close to the experimental value in benzene (37 degrees)
\cite{guto}. A different water orientation with the two OH bonds
pointing towards the graphite surface is also a local minimum, but
it has a slightly lower binding energy ($\Delta E_{{\rm b}}=8.67$\,kJ/mol).
The energy difference between this minimum and the global minimum
($\sim0.14$\,kJ/mol) is so small that zero point energy effects
might as well favor the two-legged structure as the vibrationally
averaged quantum global minimum \cite{sudiarta}. The equilibrium
distance in the global minimum between the oxygen and the graphite
surface is 3.12\,$\textrm{Å}$, which is very close to the \emph{ab
initio} value (3.04\,$\textrm{Å}$) \cite{lin} and the corresponding
values in water-C$_{60}$ (3.19\,$\textrm{Å}$) \cite{rojas} and
water-benzene (experimental, 3.33\,$\textrm{Å}$) \cite{guto}. 

\begin{figure}
\psfrag{n=2}[bc][bc]{\large $n=2$}
\psfrag{n=3}[tc][tc]{\large $n=3$}
\psfrag{n=4}[tc][tc]{\large $n=4$}
\psfrag{n=5}[tc][tc]{\large $n=5$}
\psfrag{n=6}[tc][tc]{\large $n=6$}
\psfrag{n=7}[tc][tc]{\large $n=7$}
\psfrag{n=8}[tc][tc]{\large $n=8$}
\psfrag{n=9}[tc][tc]{\large $n=9$}
\psfrag{n=10}[tc][tc]{\large $n=10$}
\psfrag{n=11}[tc][tc]{\large $n=11$}
\psfrag{n=12}[tc][tc]{\large $n=12$}
\psfrag{n=13}[tc][tc]{\large $n=13$}
\psfrag{n=14}[tc][tc]{\large $n=14$}
\psfrag{n=15}[tc][tc]{\large $n=15$}
\psfrag{n=16}[tc][tc]{\large $n=16$}
\psfrag{n=17}[tc][tc]{\large $n=17$}
\psfrag{n=18}[tc][tc]{\large $n=18$}
\psfrag{n=19}[tc][tc]{\large $n=19$}
\psfrag{n=20}[tc][tc]{\large $n=20$}
\psfrag{n=21}[tc][tc]{\large $n=21$}
\includegraphics[height=20cm,keepaspectratio]{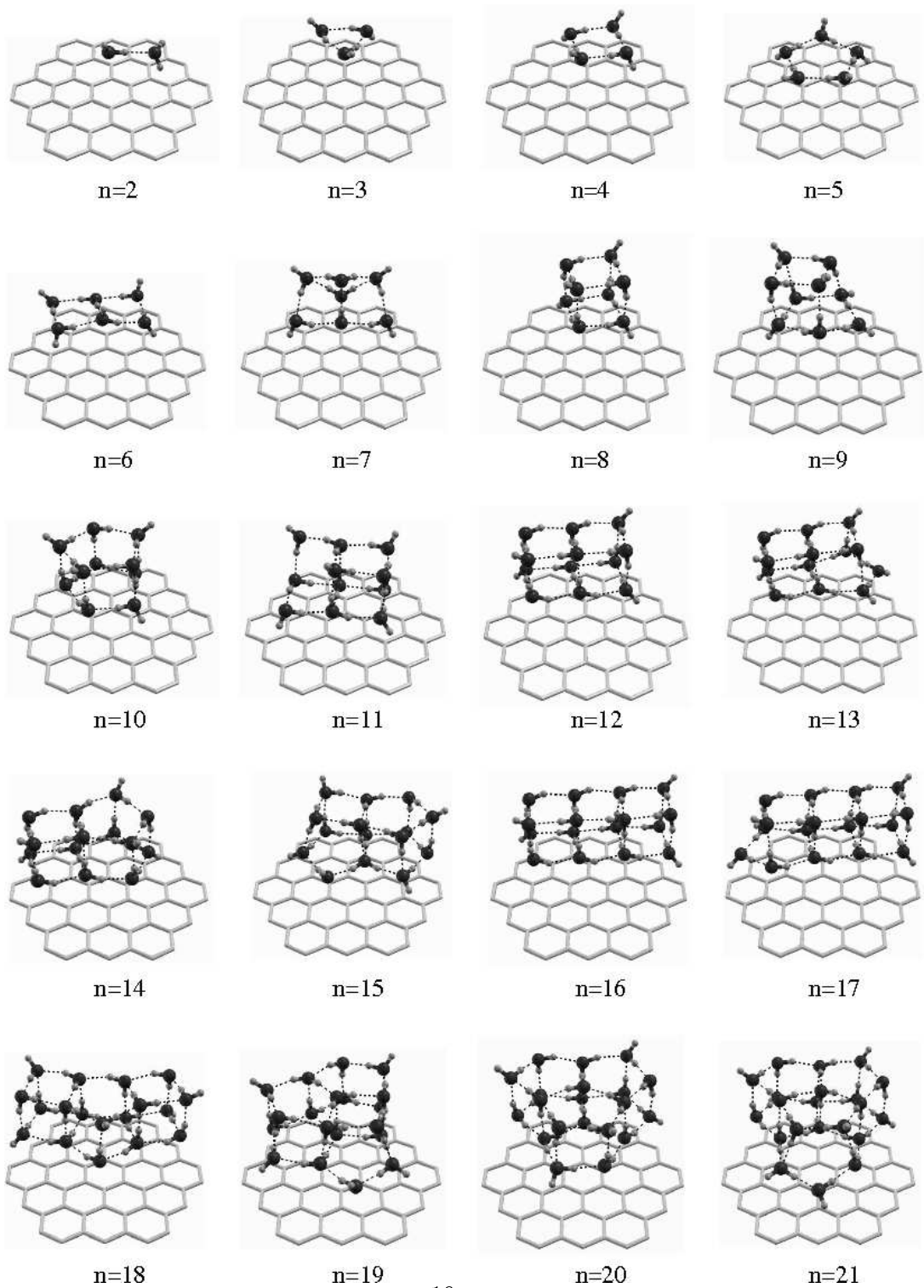}

\caption{\label{f4} Likely global minima obtained for graphite-(H$_{2}$O)$_{n}$
clusters. }
\end{figure}

The structures of the TIP4P lowest minima obtained for graphite-(H$_{2}$O)$_{n}$
are presented in Fig. \ref{f4}.%

Due to the hydrophobic nature of the water-graphite interaction, the
water substructure is often very similar to that in the corresponding
global minimum of TIP4P (H$_{2}$O)$_{n}$ \cite{wales0,kabrede}.
In some cases the structures are actually identical (aside from minor
differences in angles and distances). The exceptions are labeled in
Fig.~\ref{f2}: for those indicated by an asterisk ($n=6,\,7,\,11,\,15,\,17,\,19,\,21$)
the water substructure corresponds to a low-lying local minimum of
TIP4P (H$_{2}$O)$_{n}$, rather than the global minimum. The energy
penalty for this choice is mainly compensated by a more favorable
dispersion-repulsion contribution to the interaction energy with graphite,
which arises from a larger water-graphite contact area. For example,
in the graphite-(H$_{2}$O)$_{6}$ global minimum the geometry of
the water moiety corresponds to the {}``book'' structure, which
has also been identified as the lowest minimum in the corresponding
water-C$_{60}$ cluster \cite{rojas}. For the sizes in question,
there also exists a higher energy local minimum in which the water
substructure corresponds to the global minimum for TIP4P (H$_{2}$O)$_{n}$.
The difference between the association energies of the global and
local minimum structures of each of these graphite-(H$_{2}$O)$_{n}$
clusters is 6.36, 4.11, 1.62, 1.66, 13.04, 16.39 and 9.41\,kJ/mol
for $n=6,\,7,\,11,\,15,\,17,\,19,\,\textrm{and}\,21$, respectively.

For the clusters labeled with a circle in Fig.~\ref{f2} ($n=9,\,12,\,13$)
the water substructure corresponds to a perturbation of the TIP4P
(H$_{2}$O)$_{n}$ global minimum. In other words, relaxing the water
moiety in the absence of graphite does not lead to a nearby local
minimum, in contrast to the cases above. These graphite-(H$_{2}$O)$_{n}$
structures appear to be favored by polarization contribution in the
case $n=12$ (only the hydrogen-bond pattern is modified), and both
dispersion and polarization contributions for the other two cases.

Notice that the structure of the water moiety up to $n=6$ are basically
planar with an average oxygen-graphite distance $\overline{z}_{{\rm O}}=3.22$
$\textrm{\AA}$; the $n=7$ cluster corresponds to a transition to
two-layer water clusters ($\overline{z}_{{\rm O}}=4.04$ $\textrm{\AA}$),
and in the range $8\leq n\leq21$, we have always two-layer water
structures ($\overline{z}_{{\rm O}}=4.53$ $\textrm{\AA}$) in which
the clusters with odd $n$ have one more water molecule in the layer
closer to the graphite surface. 

The complete two-layer water structures for even $n$ are precisely
the structures of the global TIP4P free water clusters. Therefore,
these structures interact with graphite in an optimal way and they
keep their structure in the corresponding water-graphite clusters.
On the other hand, for odd $n$, the free water global minima do not
show optimal surfaces for its interaction with graphite, thus explaining
why these clusters change their structure to minimize that interaction
energy. The chosen new structures are sensibly determined by those
of either the $n-1$ or $n+1$ clusters. Other water-water potential
models do not produce this alternating behavior in the structure of
the free water global minima and, therefore, we can expect also different
behavior in the water-graphite global minima for $n\geq8$, as some
preliminary results with the TIP3P model seem to confirm. 

\begin{figure}
\psfrag{n}[tc][tc]{$n$}
\psfrag{2}[tc][tc]{$2$}
\psfrag{3}[tc][tc]{$$}
\psfrag{4}[tc][tc]{$4$}
\psfrag{5}[tc][tc]{$$}
\psfrag{6}[tc][tc]{$6$}
\psfrag{7}[tc][tc]{$$}
\psfrag{8}[tc][tc]{$8$}
\psfrag{9}[tc][tc]{$$}
\psfrag{10}[tc][tc]{$10$}
\psfrag{11}[tc][tc]{$$}
\psfrag{12}[tc][tc]{$12$}
\psfrag{13}[tc][tc]{$$}
\psfrag{14}[tc][tc]{$14$}
\psfrag{15}[tc][tc]{$$}
\psfrag{16}[tc][tc]{$16$}
\psfrag{17}[tc][tc]{$$}
\psfrag{18}[tc][tc]{$18$}
\psfrag{19}[tc][tc]{$$}
\psfrag{20}[tc][tc]{$20$}
\psfrag{-9}[br][br]{$$}
\psfrag{-8}[br][br]{-$8$}
\psfrag{-7}[br][br]{$$}
\psfrag{-6}[br][br]{-$6$}
\psfrag{-5}[br][br]{$$}
\psfrag{-4}[br][br]{-$4$}
\psfrag{-3}[br][br]{$$}
\psfrag{-2}[br][br]{-$2$}
\psfrag{-1}[br][br]{$$}
\psfrag{0}[tr][tr]{$0$}
\psfrag{1}[tr][tr]{$$}
\psfrag{2}[tr][tr]{$2$}
\psfrag{3}[tr][tr]{$$}
\psfrag{4}[tr][tr]{$4$}
\psfrag{Energy}[bl][br]{\hspace*{-2.0cm} Energy (kJ/mol)}
\includegraphics[width=8.25cm,keepaspectratio]{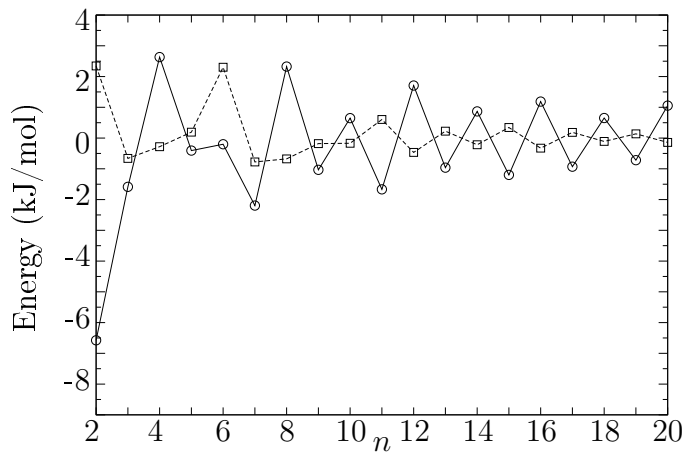}

\caption{\label{f5} Second energy differences per water molecule for the
association energies (full line) and binding energies (dashed line)
of water-graphite clusters.}
\end{figure}

The alternating behavior found in the structures of the water-graphite
global minima determines the behavior of the second energy differences.
These account for the relative cluster stability and their values
for association and binding energies, per water molecule, are plotted
in Fig. \ref{f5}. %
For $n>7$, we observe an oscillation of period $\Delta n=2$, which
corresponds to the discussed even-odd alternating structures. The
data for the association energies indicate that clusters with even
$n$ are more stable than their neighbors. This pattern is in complete
correlation with the corresponding behavior of the TIP4P free water
global minima. Fig. \ref{f5} shows that, in general, second differences
for binding and association energies are anticorrelated. The increased
stability seen in the binding energies of odd $n$ clusters relative
to their neighbors is due to their extra water molecule in close contact
with the graphite surface. From these features we conclude that the
water-water interaction certainly dominates the observed relative
cluster stability. Particularly stable clusters occur for $n=4,8,12,16,20$.

In the light of the preceding results, one could ask a couple of questions
that are relevant to assert the hydrophobic nature of the water-graphite
interaction: Are there for $n\geq7$ water-graphite local minima structures,
close in energy to the global minima, in which the water molecules
grow into a single layer (wetting structures)? How much larger should
the water-graphite interaction be for the previous wetting structures
to become the most stable ones? Answering to the second question,
we have found that we had to multiply the value of the Lennard-Jones
parameter $\varepsilon_{{\rm CO}}$ by almost a factor of two for
a wetting structure to become the global minimum. This is consistent
with the analysis performed by Werder \emph{et al}. \cite{werder}
for the monomer binding energy required to produce a wetting behavior.
By relaxing the wetting structures found by this procedure to the
closest local minimum of our original potential we have found that
these wetting local minima lie $\sim2.2$ kJ/mol per water molecule
above the global minimum. Although these values are smaller than those
found in C$_{60}$(H$_{2}$O)$_{n}$ clusters \cite{rojas}, the hydrophobic
nature of the water-graphite interaction is also a quite robust property
that would require unphysical changes in our model potential to modify
it. 

We have already shown by making use of the TIP3P potential that the
structures of the global minima of the first six water-graphite clusters
are going to depend weakly on the model chosen for the water-water
interaction. However the dependence found in the structure of the
water-graphite global minima on the structure of the corresponding
free water clusters and the known dependence of the latter on the
water-water interaction model for $n>6$, would imply changes in the
structure of these larger water-graphite clusters when a different
water model is chosen. Preliminary results confirm this prediction,
with changes that are more significant than those found for C$_{60}$(H$_{2}$O)$_{n}$
clusters.

\section{Conclusions}

\label{sec4}Using a theoretically guided empirical potential energy
surface and basin-hopping global optimization we have characterized
the geometrical structures and energetics of likely candidates for
the global potential energy minima of graphite-(H$_{2}$O)$_{n}$
clusters up to $n=21$. The structures of these minima for $1<n\leq6$
coincide with those provided by other available calculations. For
$n>2$, association energies are dominated by the water-water interaction
while the main contribution to the binding energies comes from the
dispersion energy. Our potential energy surface provides a rather
hydrophobic water-graphite interaction at the nanoscopic level. As
a consequence of this property the water substructure in the lowest
energy clusters often corresponds closely to a low-lying minimum of
the appropriate (H$_{2}$O)$_{n}$ cluster. In most cases the structure
is simply a slightly relaxed version of the global minimum for (H$_{2}$O)$_{n}$.
However, the presence of graphite can induce changes in geometry of
the water moiety.

For $n=6,\,7,\,11,\,15,\,17,\,19,\,21$ the water substructure is
based on a local minimum of (H$_{2}$O)$_{n}$, which is close in
energy to the global minimum. The energy penalty for this choice is
mainly compensated by the dispersion-repulsion contribution to the
interaction energy, because the change in structure gives rise to
a larger water-graphite contact surface. For $n=12$ the water substructure
is based on a deformation of the free (H$_{2}$O)$_{12}$ with the
same oxygen framework as the global minimum, but a different hydrogen-bonding
pattern. The different orientation of some of the OH bonds close to
the graphite surface increases the polarization energy, which stabilizes
the structure. Finally, for $n=9$ and 13 the water substructure involves
a more significant deformation of the global minimum, the energy penalty
being compensated by both the polarization and dispersion terms. A
clear alternating behavior in which the water moiety of the clusters
with even $n$ keep the structure of the corresponding free water
clusters, while their odd $n$ neighbors change it, has been observed;
this behavior has been shown to be completely correlated with that
of the free water clusters, namely (H$_{2}$O)$_{n}$ clusters with
even $n>6$ present faces that interact with graphite in an optimal
way, while those with odd $n$ do not show this feature.

Our potential energy surface also supports, for $n>6$, wetting local
minima in which the water molecules grow into a single layer. The
potential energies of these structures lie at least $2.2$\,kJ/mol
per water molecule above the global minima. The presence of the graphite
surface is necessary to stabilize the monolayer which otherwise collapses
in the cases we have considered. In this respect, we have also found
that in order to make the wetting local minima to become the cluster
global minima we should increase the magnitude of the graphite-water
interaction to unphysically high values. This implies that the hydrophobic
nature of the water-graphite interaction at very low temperature is
a quite robust property that would require unphysical changes in our
model potential to modify it.

In order to study the dependence of our qualitative picture on the
model chosen for the water-water interaction (the TIP4P potential
was our primary choice), we have repeated our calculations for the
TIP3P potential for $n\leq6$. The global minimum structures found
for these clusters coincide with those of the TIP4P model. However
the dependence found in the structure of the water-graphite global
minima on the structure of the corresponding free water clusters and
the known dependence of the latter on the water-water interaction
model for $n>6$, would imply changes in the structure of these larger
water-graphite clusters when a different water model is chosen. Preliminary
results the TIP3P model potential confirm this prediction.

The lowest energy structures obtained in the present work will be
made available for download from the Cambridge Cluster Database \cite{data}.

\section*{Acknowledgments}

This work was supported by `Ministerio de Educaci\'{o}n y Ciencia
(Spain)' and `FEDER fund (EU)' under contract No. FIS2005-02886. One
of us (BSG) also aknowledges `Ministerio de Educaci\'{o}n y Ciencia
(Spain)' for an FPU fellowship, Dr. González de Sande for his help
with code parallelization, and Cambridge University Chemical Laboratories
for the use of their computational facilitites. 
We thank Dr. D. J. Wales for his comments on the manuscript.

\end{document}